\documentclass{article}
\usepackage{amssymb}

\usepackage{graphicx}
\usepackage{amsmath}

\begin{document}

\title{{Paraconductivity in Carbon Nanotubes}}
\author{D.V.Livanov$^{1}$, A.A.Varlamov$^{2}$ \\
%EndAName
$^{1}$Moscow State Institute of Steel and Alloys (Technological University),%
\\
Leninski prospect 4, 119991 Moscow, Russia\\
$^{2}$Istituto Nazionale per la Fisica della Materia (INFM)\\
UdR ''Tor Vergata'', via del Politecnico, 1, 00133 Roma, Italy}
\maketitle

\begin{abstract}
We report the calculation of paraconductivity in carbon nanotubes above the
superconducting transition temperature. The complex behavior of
paraconductivity depending upon the tube radius, temperature and magnetic
field strength is analyzed. The results are qualitatively compared with
recent experimental observations in carbon nanotubes of an inherent
transition to the superconducting state and pronounced thermodynamic
fluctuations above $T_{c}$. The application of our results to single-wall
and multi-wall carbon nanotubes as well as ropes of nanotubes is discussed.
\end{abstract}

Carbon nanotubes are mesoscopic systems with a remarkable
interplay between dimensionality, interaction and disorder
\cite{Dek99}. Recent experiments found that the electron transport
in single-wall nanotubes (SWNT) has a one-dimensional ballistic
behavior \cite{BCLR99}. Therefore it may be theoretically
described within the model of one-dimensional interacting electron
systems known as Luttinger liquid \cite{EG97,KBF97,E99}. At the
same time, the multiwall nanotubes (MWNT), which are composed of
several concentrically arranged graphite shells, show properties
which are consistent with the weak-localization features of the
diffusive transport in magnetoconductivity and zero-bias anomaly
in the tunneling density of states \cite{BSSB99}. Similar
properties have been observed in ropes of SWNTs
\cite{BCLR99,RKKD00}.

Very recent experimental works \cite{TZWZ01,KKGR01} have addressed the
problem of superconductivity in carbon nanotubes. In the article by Tang
\textit{et al.} \cite{TZWZ01} a superconducting behavior was detected in
SWNT at a mean-field critical temperature evaluated as $T_{c}$=15 K. At the
same time a pure superconducting state with zero resistance was not found
and the authors attribute this fact to the presence of strong fluctuations
which alter severely the superconducting order parameter both below and
above $T_{c}$. In Ref. \cite{KKGR01} ropes of SWNT were studied and a truly
superconducting transition was discovered at $T_{c}$=0.55 K. The suppression
of $T_{c}$ by a magnetic field applied along the tube was also measured.

In the present communication we study the paraconductivity and corresponding
magnetoconductivity of a carbon nanotube, i.e. the contribution to the
conductivity induced by the fluctuations of the superconducting order
parameter above $T_{c}$. In what follows we assume the validity of the
Ginzburg-Landau (GL) formalism for the description of the fluctuation
superconductivity. In order to describe the one-electron spectrum of carbon
nanotubes one has to take into consideration that the electron wavelength
around the circumference of a nanotube is quantized due to the periodic
boundary conditions and only a discrete number of wavelengths can fit around
the tube. Along the tube, however, electronic states are not confined and
electrons can move ballistically. Because of the circumferential modes
quantization, the electron states in the tube do not form a single wide
energy band but instead they split into a number of one-dimensional
sub-bands with band onsets at different energies. Consequently, we assume
the electron spectrum in the form:

\begin{equation}
\epsilon (\mathbf{p})=\frac{p_{||}^{2}}{2m_{||}}+\frac{n^{2}}{2m_{\perp
}R^{2}},  \label{spectrum}
\end{equation}
where $n=-N,...,N$, $N=[p_{F}R]$, $p_{F}$ is the Fermi momentum
and $R$ is the nanotube radius. The number $N$\ is determined by
the value of the chemical potential \ and the distance between the
levels. It defines the number of electrons filling the $2N+1$
electron sub-bands of the nanotube electron spectrum. A typical
value for a realistic nanotube is $N\sim 5$ \cite{EG01}.

The longitudinal coordinate $z$ and the angular variable $\varphi $ are
chosen as the natural coordinate system for the problem under discussion.
The linearized time-dependent GL equation (TDGL) for the fluctuation order
parameter $\Psi \left( z,\varphi ,t\right) $ takes the form:

\begin{equation}
-\gamma \frac{\partial \Psi \left( z,\varphi ,t\right) }{\partial t}=%
\widehat{\mathcal{H}}\Psi \left( z,\varphi ,t\right) =\alpha T_{c}\left[
\epsilon -\xi _{\shortparallel }^{2}\frac{\partial ^{2}}{\partial z^{2}}-\xi
_{\perp }^{2}(\frac{1}{R}\frac{\mathbf{\partial }}{\partial \varphi }%
-2ieA_{\bot })^{2}\right] ,  \label{TDGLgen}
\end{equation}
where $\widehat{\mathcal{H}}$ is the GL Hamiltonian written for
the nanotube geometry, $\gamma=\pi\alpha/8$,
$\epsilon=(T-T_c)/T_c$, $A_{\bot }=HR=\frac{1}{eR}\frac{\Phi
}{\Phi _{0}}$ is the tangent component of the vector potential,
$\Phi _{0}=\pi /e$ is the magnetic flux quantum, $\xi
_{\shortparallel }=\left( 4m_{\shortparallel }\alpha T_{c}\right)
^{-1/2}$ and $\xi _{\bot }=\left( 4m_{\bot }\alpha T_{c}\right)
^{-1/2}$ are the longitudinal and the transversal GL coherence
lengths. The latter is supposed to be comparable with the nanotube
radius: $\xi _{\bot }$ $\sim R.$\ The fluctuation order parameter
$\Psi $ can be presented as a Fourier series
\begin{equation}
\Psi \left( z,\varphi ,t\right) =\sum_{n=-\infty }^{\infty
}\int_{-\pi /a}^{\pi /a}\frac{dq_{\shortparallel }}{2\pi }\psi
_{n}(q_{\shortparallel },t)\exp \left( -in\varphi \right) \exp
\left( -iq_{\shortparallel }z\right),
\label{op}
\end{equation}
and the TDGL equation for the Fourier component $\psi _{n}(t)$ is read as:

\begin{equation}
-\gamma \frac{\partial \psi _{n}(q_{\shortparallel },t)}{\partial t}%
=\varepsilon _{n}(q_{\shortparallel },\Phi )\psi _{n}(q_{\shortparallel
},t)=\alpha T_{c}\left[ \epsilon +\xi _{\shortparallel
}^{2}q_{\shortparallel }^{2}+\frac{\xi _{\perp }^{2}}{R^{2}}\left( n-\frac{%
2\Phi }{\Phi _{0}}\right) ^{2}\right] \psi _{n}(q_{\shortparallel },t).
\label{tdgl}
\end{equation}
Here $\varepsilon _{n}(q_{\shortparallel },\Phi )$ are the eigenvalues of $\
\widehat{\mathcal{H}}.$

The angular quantization gives rise to rather distinctive critical
temperatures corresponding to the different order parameter modes. These
critical temperatures are: $T_{c}^{\left( n\right) }(\Phi =0)=T_{c}^{\left(
0\right) }\left[ 1-\left( \frac{\xi _{\bot }^{2}}{R^{2}}\right) n^{2}\right]
$ and the characteristic dimensionless temperature difference is $\Delta
\varepsilon _{0}\sim \frac{\xi _{\bot }^{2}}{R^{2}}.$ As expected, when the
temperature decreases the system tends to the mode with $n=0$. In nonzero
magnetic fields, when the magnetic flux is $\Phi \in $ $]-\Phi _{0}/2$\ $%
,\Phi _{0}/2[$, the superconducting transition occurs at the $\psi _{0}$
state.

Let us move to the study of the paraconductivity in a small
superconducting cylinder at temperatures above the critical one.
The fluctuation-induced current can be expressed by its general
quantum mechanical form averaged over all possible values of the
fluctuation order parameter $\Psi \left( z,\varphi ,t\right) .$
The latter can be defined as the solution of the TDGL equation
(\ref{TDGLgen}) when the Langevin forces are introduced in the
right hand side. After some algebra the general expression for the
paraconductivity tensor \ can be obtained (see details in
\cite{LV02} ) in
the form of a convolution product of the velocity matrix elements $\widehat{%
\mathbf{v}}_{\{li\}}^{\alpha }$ and the kernel containing the eigenvalues of
the GL Hamiltonian of Eq. (\ref{TDGLgen}). We will only need its
longitudinal diagonal component:

\begin{equation}
\sigma ^{\Vert }(\epsilon ,H)=\frac{\pi \alpha e^{2}}{2}T\sum_{\{i,l\}=0}^{%
\infty }\frac{\widehat{\mathbf{v}}_{\{il\}}^{z}\widehat{\mathbf{v}}%
_{\{li\}}^{z}}{\varepsilon _{\{i\}}\varepsilon _{\{l\}}\left( \varepsilon
_{\{i\}}+\varepsilon _{\{l\}}\right) }.  \label{sym}
\end{equation}
The appropriate matrix elements of the velocity operator are

\begin{equation}
\widehat{\mathbf{v}^{z}}_{il,pq}=\mathbf{v}_{p}\delta _{pq}\delta _{il},\;%
\mathbf{v}_{p}^{z}=\frac{\partial \varepsilon _{p}}{\partial p}=2\alpha
T_{c}\xi _{\shortparallel }^{2}p.  \label{matr}
\end{equation}
The summation over subscript $\{i\}$ is carried out over the levels of the
angular quantization up to the maximal number $N$ and includes the
integration over the $z$-axis momentum. As a result the general formula (\ref
{sym}) for the longitudinal paraconductivity of a nanotube is read as

\begin{eqnarray}
\sigma ^{\shortparallel }(\epsilon ,H) &=&\frac{\pi \alpha e^{2}}{2S}T\int
\frac{dp_{\shortparallel }}{2\pi }\int \frac{dq_{\shortparallel }}{2\pi }%
\sum_{i,l=-N}^{N}\frac{\widehat{\mathbf{v}}_{il,pq}^{\shortparallel }%
\widehat{\mathbf{v}}_{li,qp}^{\shortparallel }}{\varepsilon _{i}\left(
p_{\shortparallel }\right) \varepsilon _{l}\left( q_{\shortparallel }\right) %
\left[ \varepsilon _{i}\left( p_{\shortparallel }\right) +\varepsilon
_{l}\left( q_{\shortparallel }\right) \right] }=  \label{gener} \\
&=&\frac{\pi e^{2}}{16S}\xi _{\shortparallel }\sum_{n=-N}^{N}\frac{1}{\left[
\epsilon +\frac{\xi _{\perp }^{2}}{R^{2}}\left( n-\frac{2\Phi }{\Phi _{0}}%
\right) ^{2}\right] ^{3/2}}  \notag
\end{eqnarray}
(here $S=\pi R^{2}$ is the cross-section area of the nanotube). \ This
formula can be numerically evaluated to obtain the magnetoconductivity.
Nevertheless, in order to get a qualitative understanding of the
paraconductivity temperature dependence in zero field and its behavior in
the presence of a magnetic field at fixed temperature, let us assume $N\gg 1$
and try to proceed analytically.

\textbf{1. Zero magnetic field. }The most convenient way to
analyze Eq. (\ref {gener}) is to isolate the term with $n=0$ and
treat the first term and the remaining sum separately. Relatively
far from the critical temperature, where $\xi _{\perp }(\epsilon
)=\xi_{\perp}/\sqrt{\epsilon}\ll R$, (but still $\epsilon \ll 1),$
one can replace the sum in Eq. (\ref{gener}) with an integral to
get

\begin{equation}
\sigma ^{\shortparallel }(\epsilon ,0)=\frac{\pi e^{2}}{16S}\xi
_{\shortparallel }\frac{1}{\epsilon ^{3/2}}+\frac{\pi e^{2}}{8S}\xi
_{\shortparallel }\left( \frac{R}{\xi _{\perp }}\right) ^{3}\left\{ \frac{1}{%
\sqrt{1+\frac{R^{2}}{\xi _{\perp }^{2}(\epsilon )}}\left[ 1+\sqrt{1+\frac{%
R^{2}}{\xi _{\perp }^{2}(\epsilon )}}\right] }-\frac{1}{\sqrt{N^{2}+\frac{%
R^{2}}{\xi _{\perp }^{2}(\epsilon )}}\left[ N+\sqrt{N^{2}+\frac{R^{2}}{\xi
_{\perp }^{2}(\epsilon )}}\right] }\right\} .  \label{zerofi}
\end{equation}
One can see that for temperatures far enough from $T_c$, such that
$\xi _{\perp }(\epsilon )\ll R/N$, the $1D$ limit is reached while
in the interval where $R/N\ll \xi _{\perp }(\epsilon )\ll R$ the
Eq. (\ref{zerofi}) reproduces the $2D$ result for
paraconductivity. In the immediate vicinity of the critical
temperature, where $\xi _{\perp }(\epsilon )\gg R$, \ only the
first term in Eq. (\ref {zerofi}) contributes to the
paraconductivity and the system is again in the $1D$ limit.

All the asymptotics of Eq. (\ref{zerofi}) can be presented in a more compact
form as:

\begin{equation}
\sigma ^{\shortparallel }(\epsilon ,0)=\frac{e^{2}\xi _{\shortparallel }}{%
16R^{2}}\left\{
\begin{tabular}{l}
$\frac{1}{\epsilon ^{3/2}},\;\epsilon \ll \left( \frac{\xi _{\perp }}{R}%
\right) ^{2}$ \\
$\left( \frac{R}{\xi _{\perp }}\right) \frac{2}{\epsilon },\left( \frac{\xi
_{\perp }}{R}\right) ^{2}\ll \epsilon \ll \left( \frac{\xi _{\perp }N}{R}%
\right) ^{2}$ \\
$\frac{2N+1}{\epsilon ^{3/2}},\left( \frac{\xi _{\perp }N}{R}\right) ^{2}\ll
\epsilon $%
\end{tabular}
\right. .  \label{asy}
\end{equation}

The physics of these crossovers is the following. The first one
has a geometrical nature: very near to $T_{c}$ the fluctuation
Cooper pairs are so large that they have only one degree of
freedom to slide along the tube axis. The first line of Eq.
(\ref{asy}) exactly reproduces the paraconductivity of a wire with
cross-section $S\ll $ $\xi _{\perp }^{2}$. \ In the intermediate
regime rotations over the tube surface become possible and the
paraconductivity temperature dependence transforms into the $2D$
one. Finally, relatively far from $T_{c},$ where $\xi _{\perp
}(\epsilon )\sim R/N $ , the last, most nontrivial, crossover
$2D\rightarrow 1D$ in the fluctuations dimensionality takes place.
Let us stress the longitudinal
paraconductivity in this range of temperatures acquires a degeneracy factor $%
2N+1$\ equal to the number of electron sub-bands.

\textbf{2. Non-zero magnetic field.} We now move to the study of
paraconductivity in the presence of a magnetic field applied. Due to the
Little-Parks effect \cite{Little-Parks}, the critical temperatures $%
T_{c}^{\left( n\right) }(\Phi )$ are periodic functions of the
flux through the tube with period $\Phi _{0}$. Therefore we can
restrict ourselves to the flux range $-\Phi _{0}/2$ $<\Phi $
$<\Phi
_{0}/2,$ where $T_{c}\left( \Phi \right) =T_{c}^{\left( 0\right) }\left[ 1-%
\frac{\xi _{\perp }^{2}}{R^{2}}\left( \frac{2\Phi }{\Phi _{0}}\right) ^{2}%
\right] .$ Evidently, two different regimes can take place: a
weak-field one when $\Phi \lesssim $ $\Phi _{0}\frac{R}{\xi
_{\perp }}\sqrt{\epsilon }$
(which is equivalent to $H\lesssim H_{c2}\left( \frac{R}{\xi _{\perp }}%
\right) \sqrt{\epsilon }$) and a strong-field regime when $\Phi _{0}\frac{R}{%
\xi _{\perp }}\sqrt{\epsilon }\ll \Phi \ll $ $\Phi _{0}/2.$ The general
formula (\ref{gener}) may be rewritten as

\begin{equation} \label{fieldgen}
\sigma ^{\shortparallel }(\epsilon ,\Phi )=\frac{\pi e^{2}}{16S}\xi
_{\shortparallel }\left( \frac{R}{\xi _{\perp }}\right) ^{3}\frac{1}{\left[
\left( \frac{2\Phi }{\Phi _{0}}\right) ^{2}+\frac{R^{2}}{\xi _{\perp }^{2}}%
\epsilon \right] ^{3/2}}+\frac{\pi e^{2}}{16S}\xi _{\shortparallel
}\left( \frac{R}{\xi _{\perp }}\right)
^{3}\sum_{n=1}^{N}\frac{1}{\left[ \left(
n\pm \frac{2\Phi }{\Phi _{0}}\right) ^{2}+\frac{R^{2}}{\xi _{\perp }^{2}}%
\epsilon \right] ^{3/2}}.
\end{equation}

In the case of the weak-field regime one can easily see that the
main magnetic field dependence comes from the renormalization of
the critical temperature, so

\begin{eqnarray}
\delta \sigma ^{\shortparallel }(\epsilon ,\Phi ) &=&\sigma ^{\shortparallel
}(\epsilon ,\Phi )-\sigma ^{\shortparallel }(\epsilon ,0)=  \notag \\
&=&-\frac{e^{2}\xi _{\shortparallel }}{8}\frac{\xi _{\perp }^{2}}{R^{4}}%
\left( \frac{\Phi }{\Phi _{0}}\right) ^{2}\left\{
\begin{tabular}{l}
$\frac{3}{\epsilon ^{5/2}},\;\epsilon \ll \left( \frac{\xi _{\perp }}{R}%
\right) ^{2}$ \\
$4\left( \frac{R}{\xi _{\perp }}\right) \frac{1}{\epsilon ^{2}},\left( \frac{%
\xi _{\perp }}{R}\right) ^{2}\ll \epsilon \ll \left( \frac{\xi _{\perp }N}{R}%
\right) ^{2}$ \\
$\frac{3\left( 2N+1\right) }{\epsilon ^{5/2}},\left( \frac{\xi _{\perp }N}{R}%
\right) ^{2}\ll \epsilon $%
\end{tabular}
\right.
\end{eqnarray}

The strong-field regime $\Phi _{0}\frac{R}{\xi _{\perp
}}\sqrt{\epsilon }\ll \Phi \ll $ $\Phi _{0}/2$ can be reached
(without passing to the next foil of the Little-Parks effect) in
the case $R\ll \xi_{\perp} \left( \epsilon \right) $ only. In this
case one finds that the main contribution originates from the
first term in Eq. (\ref{fieldgen}):

\begin{equation}
\sigma ^{\shortparallel }(\Phi )=\frac{e^{2}}{2^{7}}\frac{R\xi
_{\shortparallel }}{\xi _{\perp }^{3}}\left( \frac{\Phi _{0}}{\Phi }\right)
^{3}.  \label{strofield}
\end{equation}
This result is valid for temperatures $\epsilon \ll \left( \frac{\xi _{\perp
}}{R}\right) ^{2}.$ In the temperature range $\left( \frac{\xi _{\perp }}{R}%
\right) ^{2}\ll \epsilon \ll \left( \frac{\xi _{\perp }N}{R}\right) ^{2}\ll
1 $ (if such interval exists), where in the absence of the magnetic field
the fluctuations have a $2D$ character, the effect of the magnetic field is
relevant only for fields so high as $\Phi _{0}N\frac{R}{\xi _{\perp }}\sqrt{%
\epsilon }\ll \Phi \ll $ $\Phi _{0}/2$, but \ it still is described by the
formula (\ref{strofield}). One can recognize in the effect of the magnetic
field on paraconductivity the usual suppression of the effective fluctuation
dimensionality, as it happens even in the $3D$ case. Nevertheless we would
like to attract the reader's attention to the unusually strong suppression
of the nanotube paraconductivity in strong magnetic fields. Its comparison
with the corresponding paraconductivity of a layered superconductor shows a
remarkable difference in the critical exponent: 3 against 1 (see Ref. \cite
{BV98}). This follows from the channel separation in the Cooper pairs motion
and hence the effective decrease of their density in the momentum space. A
similar effect is observed in superconducting rings. In its $0D$ regime $%
\sigma _{ring}^{\left( 0\right) }(H)\sim H^{-4}$ (see Ref.
\cite{BV02}) instead of \ $\sigma _{gran}^{\left( 0\right)
}(H)\sim H^{-2}$ as for superconducting granules (see Refs.
\cite{LV02,V. V. Schmidt}).

Let us discuss the results obtained. In Fig. 1 we have plotted the
resistivity of carbon nanotubes calculated from
Eq.(\ref{fieldgen}), choosing $N=5$ for different magnetic field
strengths. It can be seen that the simulated behavior is similar
to the experimental one reported in Ref. \cite{KKGR01} for
metallic ropes of nanotubes. Discussing the application of our
results to recent experimental data concerning realistic nanotubes
\cite {BCLR99,BSSB99,TZWZ01,KKGR01}, it is important to remember
that the physics of superconductivity in these systems is still
controversial and it is very likely to be qualitatively different
for systems like MWNT, the ropes of SWNTs or individual SWNT.
Namely, the effect of interactions within multiwall tubes or
hopping between neighboring tubes in a rope drives the system away
from the one-dimensionality characterizing an individual nanotube.
Therefore the physical properties are substantially altered in
both the normal and superconducting states depending on whether
hopping is effective or not. Nevertheless our considerations are
quite general because the model proposed is based on the GL
phenomenology which is independent of the specific pairing
mechanism leading to the superconductivity. It is clear that an
individual nanotube is rather within the one-dimensional limit of
Eq. (\ref{zerofi}), $\xi _{\perp }(\epsilon )\gg R$, while for
multiwall tubes or ropes the other regimes may be observed.

As it was demonstrated above, the nontrivial geometry of tube
leads to a number of possible crossovers in the temperature
dependence of the paraconductivity. The crossover closest to the
critical temperature has a clear geometric nature and is analogous
to the one occurring in thin films of layered superconductors
\cite{VY91}. As the system moves away from the transition point,
the coherence length decreases, thus rotations over the tube
surface become possible and the system goes into the $2D$ regime.
The last $2D\rightarrow 1D$ crossover has an intrinsic origin, it occurs when $%
\xi _{\perp }(\epsilon )\lesssim R/N$.

The alternative interpretation of different regimes of
paraconductivity behavior can be given on the basis of comparison
of the characteristic fluctuation Cooper pair ''binding energy'',
$T-T_{c}$, with the angular quantization energy level structure.
Here it is necessary to remind that the fluctuation Cooper pairs
above the critical temperature are not condensed with the zero
energy, like it happens below $T_{c},$ but they are distributed
over energy with the rapid decay at $\varepsilon \gtrsim T-T_{c}.
$ When $\xi _{\perp }(\epsilon )\gg R$ ($T-T_{c}\ll 1/2m_{\perp
}R^{2})$ the binding energy is so small that the electrons
occupying only the $n=0$ level \ can be involved in fluctuation
pairing. In result the $1D$ behavior takes place. As $T-T_{c}$
growths ($R/N\lesssim \xi _{\perp }(\epsilon )\lesssim R$) \ the
electrons from more and more subbands can be involved in pairing
(within the same subband) and due to this additional degree of
freedom (subband number $n$) the fluctuation behavior becomes
$2D.$ Finally, when $T-T_{c}$ exceeds the energy of the last
filled level of angular quantization $\varepsilon _{N}=$
$N^{2}/2m_{\perp }R^{2}$( what means $\xi _{\perp }(\epsilon
)\lesssim R/N$) all $2N+1$ subbands are involved in pairing and
each one presents the independent one-dimensional channel. Indeed,
the corresponding formula differs from the one near $T_{c}$ by a
factor $2N+1$ (see Eqs. (\ref{asy}) and (\ref{fieldgen})).

The experimental observation of such crossovers, side by side with the
strong suppression of the paraconductivity by magnetic fields would be good
''pro'' or ''contra'' arguments in the discussion of the validity of the GL
phenomenological approach to the study of superconductivity in such
nontrivial objects as nanotubes.

The authors are grateful to B.L.Altshuler and G.Balestrino for useful
discussions and acknowledge the financial support of the NATO Collaborative
Linkage Grant \ CLG \ 978153, PA TIN INFM, Tor Vergata - MISA Collaboration
Program, Program ''New Materials'' of the Ministry of Education of Russia.

\newpage

\begin{center}
\textbf{Figure caption}
\end{center}

Theoretical prediction for the temperature dependence of the
resistivity of carbon nanotube. Plotted is the resistivity
calculated as a sum of normal-state temperature-independent
contribution and paraconductivity versus the reduced
temperature $(T-T_c(0))/T_c(0)$. The field strengths are $\Phi/\Phi_0=0,$ $%
0.25$ and $0.5$.

\end{document}